## Title:

**Tuning the secondary electron yield and electronic conductivity of $Mg_XZn_{(1-x)}O$ multilayers synthesized by Atomic layer deposition.**

## Authors:

**M. Lafarie[1,2], S. Dadouch[2], F. Miserque[3], Y. Zheng[4], J. Leroy[5], M. Belhaj[3], T. Proslier[1]**

**1) IRFU, CEA Université Paris-Saclay, Gif Sur Yvette, France**

**2) DPHY, ONERA, Toulouse, France**

**3) LECA, CEA Université Paris-Saclay, Gif Sur Yvette, France**

**4) INSP, CNRS Université Paris Sorbonne, Paris, France**

## Abstract

Secondary electron emission (SEE), arising from the interaction of energetic electrons with material surfaces, can induce deleterious effects in vacuum radio-frequency (RF) systems, notably multipacting, thereby degrading performance and threatening device integrity. Despite extensive efforts, effective and widely adopted mitigation strategies remain limited. Here, we report an approach to simultaneously tailor SEE and electrical conductivity using nanometric multi-element multilayers deposited by atomic layer deposition (ALD). The design and fabrication of these heterostructures are presented, and a range of films is systematically characterized to establish correlations between emission properties, electrical conductivity, chemical and structural features. We demonstrate that such heterostructures enable combinations of total electron emission yield (TEEY) and conductivity unattainable in single-phase materials, including regimes combining high TEEY with relatively high conductivity and low TEEY with low conductivities. These results open new avenues for the rational design of functional surfaces for vacuum electronic applications.

## Highlights

- Successful ALD growth of ZnMgO films with wide difference in composition and structure
- Correlation established between ALD growth parameters, ZnMgO films chemical composition and morphology, TEEY and electrical conductivity
- Successful demonstration of TEEY / electrical conductivity independent modulation

## Introduction

The total electron emission yield (TEEY) refers to the number of electrons emitted (secondaries and backscattered) from a surface for each incident electron of a given energy. TEEY is possibly at the origin of two problematic phenomena: electrostatic discharges (ESD) on satellites and multipacting in vacuum radiofrequency (RF) components.

ESDs on satellites particularly affect their solar panels. In the case of a satellite in low Earth orbit (LEO), its metallic structure and its solar cells, including the dielectric coverglasses, share the same potential, around –3 kV (this potential varies with orbital altitude; in geostationary orbit (GEO) it can reach up to –25 kV [1]). When the satellite is exposed to sunlight, it emits electrons through photoemission and interactions with charged particles from the solar wind. Consequently, the sunlit areas lose previously accumulated negative charges and their electrical potential increases. However, since the structure is generally metallic and therefore conductive, the potential can be equalized, preventing the formation of a large potential difference between the illuminated and non-illuminated sides. Overall, the potential remains around –3 kV. The exception comes from the dielectrics, mainly the coverglasses, which represent the largest exposed dielectric surface. Their potential rises due to electron emission (photoemission and TEEY > 1) as long as they are exposed to sunlight. As a result, a potential difference of up to 1500 V [2] can develop between the coverglass surfaces and the rest of the satellite. Such a difference can trigger ESDs, creating a high risk of surface damage to the solar panels and therefore a reduction of the available electrical power for the satellite. The solutions proposed to mitigate the occurrence of ESDs have mainly focused on the electrical design of solar panels [1, 3].

Multipacting is a major issue for RF systems in space applications. Indeed, the pursuit of wider RF bands toward higher frequencies and the increase in RF power make multipacting prevention more complex. Moreover, multipacting generated on dielectric surfaces can lead to a flashover, which may permanently damage the affected components and thus compromise the operation of the entire system. This risk is therefore incompatible with the level of reliability and resilience required for space applications. Filters [4] and coaxial transmission lines [2] are among the on-board RF components at risk. The same phenomenon is found in particle accelerators, where superconducting radio-frequency (SRF) resonant cavities are commonly used to accelerate charged particles. The particle energy gain is directly controlled by raising the RF power and the accelerating gradient of a TE01 mode inside the SRF cavity. Operating at higher gradients however, can trigger a sudden degradation of the cavity quality factor, mainly as a result of multipacting.

In the field of vacuum radio-frequency (RF) components in general, increasing RF-power can lead to an abrupt drop in their performance, due to multipacting. The origin of multipacting lies in the presence in these systems of free electrons [5], extracted from the surface of the system by field emission or by interaction with cosmic particles. These electrons interact with the RF field used in the vacuum system. In particular, they can be deflected and accelerated close to their point of extraction, or onto an opposite surface. However, if the surface has a total electron emission efficiency (TEEY) greater than 1 for the impact energy of these electrons (which reaches between 30 eV and 200 eV in cubic copper or niobium resonant cavities [6]), a self-sustaining electron cloud can form within a few RF periods. This cloud will then absorb an increasing proportion of the RF power and a growing number of electron impacts may locally heat the device surface, eventually causing the material to lose its superconducting properties—a process known as quenching for SRF cavities [7,8,9].

Conversely, high electron emission yields are desirable in microchannel plate (MCP) electron multipliers, which amplify weak signals through successive electron impacts on the inner walls of microscopic channels under an applied electric field. A high TEEY, dominated by secondary electron emission at the relevant impact energies, promotes electron multiplication and enhances detector gain. The channel walls must also exhibit carefully controlled electrical conductivity to replenish the emitted charge and prevent local charging and gain saturation, while limiting dark current and Joule heating. Independently tuning electron emission and charge transport is therefore essential to balance amplification, recovery time, and rate capability. ALD offers a powerful route to this control by conformally coating the channels with resistive and emissive layers whose composition and thickness can be tailored separately, enabling high secondary electron yields alongside an optimized electrical resistance.

Atomic Layer Deposition (ALD) have been previously successfully used to mitigate multipacting in SRF cavities [10,11]. Thanks to the atomic-level thickness control offered by ALD, the electrical conductivity of a TiN layer could be precisely adjusted while maintaining a low TEEY. This strategy successfully removed the multipacting barrier while preserving quality factors compatible with operation in a particle accelerator of $10^{10}$. These initial results highlighted ALD's ability to precisely control film thickness. however, only a single coating alloy was investigated and the only tuning one parameter, the layer thickness, did not allow for an independent tuning of the TEEY and conductivity. For the applications mentioned previously, a large palette of TEEY and conductivity (i.e insulator) values have to be independently achievable within one material; from low TEEY and low conductivity to a high TEEY and high conductivity. Such combinations are impossible with single component bulk materials and novel heterostructure have to be designed.

In this work we show how the TEEY and electrical conductivity can be adjusted independently by controlling precisely the structure end chemical composition of a two-material, namely MgO and ZnO, heterostructures.

## Experimental details

The two materials selected in this study are ZnO and MgO because their TEEY (2 [12] for ZnO and 6.2 [13] for MgO) and electrical conductivity (1.105 $\Omega^{-1}.m^{-1}$ for ZnO [14] and $10^{-15}$ $\Omega^{-1}.m^{-1}$ for MgO [15]) values are very different and their separate and combined synthesis [16] is already detailed in the literature.
A homemade flux flow ALD bench at CEA Saclay was used to synthesize the heterostructures. It operates at a pressure of around 1.1 mbar and under laminar flow of $N_2$ (300 sccm during growth) [11]. A residual gas analyzer was used in-situ to monitor ALD processes during the MgO and ZnO synthesis. films were grown at 150°C. ZnO was synthesized from DiEthylZinc (DEZ) and $H_2O$ and MgO with $Mg(EtCp)_2$ and $H_2O$. The ALD cycles parameters used to deposit ZnO and MgO are 1.5 s pulse / 15 s $N_2$ purge for all precursors. The ZnMgO heterostructures were deposited on a boron-doped Si (111) substrate with a $SiO_2$ thermal oxide layer (100nm thick) on which was deposited 200 ALD cycles (approximately 20nm) of alumina (2 s TMA / 10 s $N_2$ / 1.5 s $H_2O$ / 10 s $N_2$) at 300°C.
Heterostructure thickness, structure and density were measured by X-Ray reflectivity (XRR) and Grazing Incidence X-Ray Diffraction (GIXRD) (Rigaku SmartLab, monochromated Cu Kalpha line) using X'pert reflectivity function software with the Parrat formalism [17, 18] and CristalDiffract@ software to fit the experimental data.
X-ray Photoelectron Spectroscopy (XPS), TEEY, conditioning and surface erosion analyses were carried out on ONERA's ALCHIMIE bench [19] operating at pressures between $2.10^{-9}$ mbar and $7.10^{-10}$ mbar. XPS measurements were carried out with an X-ray source allowing the use of the non-monochromated Kalpha line of Al (1486.6 eV) or Mg (1253 eV). The Al Kalpha line was systematically used, with an anode bias of 15 kV and an emission current of 20 mA. Each sample was analyzed in survey, with a CAE of 100 eV, a dwell time of 0.3 s and averaged over 25 runs. Then a high-resolution spectrum was acquired for the $Mg_{1s}$, $Mg_{2p}$, $Zn_{2p\ 3/2}$, $Zn_{3s}$, $O_{1s}$ and $C_{1s}$ peaks with a dwell time of 0.5 s, an EAC of 20 eV, an energy step of 0.1 eV and averaged over runs.
High-resolution XPS spectra were analyzed using Casa XPS software based on literature data ($C_{1s}$ [20, 21, 22, 23]; $Mg_{1s}$ [23] and $Mg_{2p}$ [23, 24]; $Zn_{2p}$ [23, 25] and $Zn_{3s}$ [23]; $O_{1s}$ [20, 23, 25, 26, 27]).
Surface conditioning was carried out with an e- gun (Kimball physics 1 keV Flood Gun) with electron energy set at 500 eV with an irradiation current of about 7 $nA.mm^{-2}$ with a 45° incidence angle for at least 14 h.
Surface erosion was performed with a source (FOCUS) under Ar+ ion flux at 1keV for 210 to 240 s with an emission current of around 2 µA. These parameters were chosen because XPS analyses showed a clear reduction in carbon content and that erosion of the coatings themselves was negligible after comparison of the survey composition between a first erosion and a second erosion stage.
The TEEY measurements were carried out using an $e^-$ gun (Kimball Physics 1 eV - 2 keV) with output energies ranging from 10 to 2,000 eV, coupled to a function generator which enabled the measurements to be performed in pulsed electron beam mode. TEEY measurements were performed using the sample current method [28].
In addition, before both the incident electron current, $I_0$ and the sample current, $I_s$ measurements, samples were irradiated with the X-ray source used for XPS measurements for 10 minutes. All other parameters being equal, irradiation does not alter the measured TEEY for samples that are sufficiently conductive not to accumulate space charges during TEEY measurements.
The electrical conductivity of heterostructures was measured using the 4-point method at room temperature and in ambient atmosphere.

## Results

### a) Microstructural and chemical characterization

In order to investgate the influence of the chemical and structral properties on the thin films TEEY and electrical conductivity, 18 ZnMgO heterostructures, 1 pure ZnO sample and 1 MgO sample were produced. Table 1 summarizes the ALD cycles and supercycles associated with each deposit.

Table 1: Summary of the different samples presented in this study

| Nomenclature | Supercycle description | Supercycles | Surface layer (ALD cycles) |
|---|---|---|---|
| ZnO | 200 cycles of ZnO | 1 | ZnO |
| 1/16* | 1 cycle of MgO + 16 cycles of ZnO | 12 | ZnO (16) |
| 1/8* | 1 cycle of MgO + 8 cycles of MgO | 24 | ZnO (8) |
| 1/4* | 1 cycle of MgO + 4 cycles of ZnO | 42 | ZnO (4) |
| 1/2* | 1 cycle of MgO + 2 cycles of ZnO | 72 | ZnO (2) |
| 1/1* | 1 cycle of MgO + 1 cycle of ZnO | 112 | ZnO (1) |
| 2/2* | 2 cycles of MgO + 2 cycles of ZnO | 56 | ZnO (2) |
| 4/4* | 4 cycles of MgO + 4 cycles of ZnO | 28 | ZnO (4) |
| 8/8* | 8 cycles of MgO + 8 cycles of ZnO | 14 | ZnO (8) |
| 16/16* | 16 cycles of MgO + 16 cycles of ZnO | 7 | ZnO (16) |
| 32*/32 | 32 cycles of ZnO + 32 cycles of MgO | 6 | MgO (32) |
| 16*/16 | 16 cycles of ZnO + 16 cycles of MgO | 7 | MgO (16) |
| 12*/12 | 12 cycles of ZnO + 12 cycles of MgO | 9 | MgO (12) |
| 8*/8 | 8 cycles of ZnO + 8 cycles of MgO | 14 | MgO (8) |
| 4*/4 | 4 cycles of ZnO + 4 cycles of MgO | 28 | MgO (4) |
| 1*/1 | 1 cycle of ZnO + 1 cycles of MgO | 112 | MgO (1) |
| 1*/2 | 1 cycle of ZnO + 2 cycles of MgO | 77 | MgO (2) |
| 1*/4 | 1 cycle of + 4 cycles of MgO | 48 | MgO (4) |
| 1*/8 | 1 cycle of + 8 cycles of MgO | 28 | MgO (8) |
| 1*/16 | 1 cycle of ZnO + 16 cycles of MgO | 15 | MgO (16) |
| MgO | 250 cycles of MgO | 1 | MgO |

The chemical composition of the heterostructures in series 1 and 2 was analyzed in their "as received" surface states, after electronic conditioning, and after ion erosion. Figure 1 shows the Zn ratio measured for the different heterostructures. What is referred to as the "Zn ratio" in the rest of the article is the ratio of the atomic percentage of Zn to the sum of the atomic percentages of Zn and Mg measured by XPS. The Zn ratio is therefore calculated as follows

$$Zn\ Ratio\ = \frac{\%at\ Zn}{\%at\ (Zn + \%at\ Mg)}\ (1)$$

Where :

- $\%_{at}$ Zn and $\%_{at}$ Mg are the atomic percentage of respectively Zn and Mg obtained by analyzing the XPS spectrum survey of the corresponding sample.

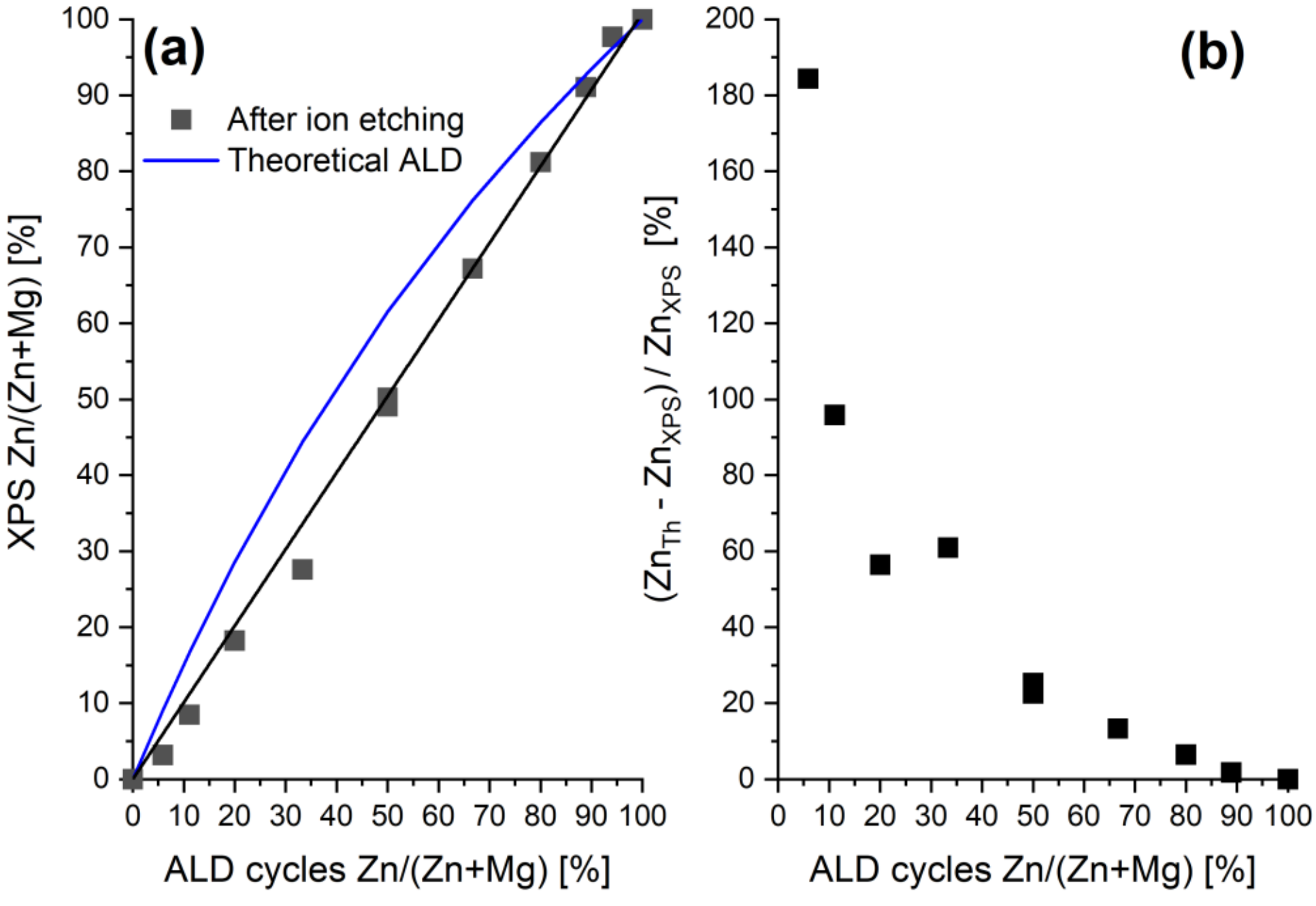

Figure 1: Graphs showing (a) the Zn ratio of ZnMgO heterostructures and (b) the corresponding relative difference between theoretical and measured Zn ratios as a function of the ZnO ALD cycle ratio for ZnMgO heterostructures in series 1. The relative error bars for each measurement are combined in the thickness of the dots.

The Zn ratio measured after etching is considered representative of the intrinsic composition of the heterostructures. Indeed, etching reduces the initially measured surface carbon contamination by more than 80% as compared to the as-received state, lowering it to only 1–4 at.%.

The Zn ratio for series 1 heterostructures is plotted in figure 1a as a function of the ratio of ZnO ALD cycles, representing the targeted composition. The intrinsic Zn ratio matches very well the intended ZnO ALD cycles carried out for the synthesis of the heterostructures and can be approximated by a linear function of the ratio of ZnO cycles performed, as represented by the linear fit (black curve) in Figure 1a.

The theoretical Zn ratio curve shown in figure 1a (doted curve) is obtained as follows:

$$Theoretical\ Zn\ Ratio = \frac{ZnO\ cycles\ \times ref\ ZnO\ GPC}{ZnO\ cycles \times\ ref\ ZnO\ GPC + MgO\ cycles \times\ ref\ MgO\ GPC} \quad (2)$$

Where :

- ZnO cycles is the number of ZnO ALD cycles performed for a given heterostructure.
- MgO cycles is the number of MgO ALD cycles performed for a given heterostructure.
- Reference growth per cycle (GPC) for ZnO = 1.6-1.7 Å.cycle$^{-1}$, obtained from previous measurement on pure ALD ZnO film is in agreement with the literature [29].
- Reference GPC for MgO = 1.2-1.3 Å.cycle$^{-1}$, obtained from previous measurement on pure ALD MgO film consistent with the value usually reported in the literature [30].

The theoretical Zn ratio is higher than the Zn ratio measured by XPS after etching for Zn ratio ≤ 80%, indicating that the when MgO is present in the film, the ZnO growth rate per cycle (GPC) is different from the value obtained for pure ZnO and the literature reference.

This trend is better illustrated in figure 1b that presents the variation in the relative difference between the theoretical and experimental Zn ratios for the series 1 heterostructures, excluding pure ZnO and pure MgO samples. It shows that as the number of ZnO ALD cycles increases, the measured Zn ratio approaches the theoretical value. Conversely, when more MgO cycles are used, the deviation from the theoretical Zn ratio becomes significant. This trend suggests that in MgO-rich films, the growth rate of ZnO tends to decrease and progressively shifts toward the growth behavior characteristic of MgO.

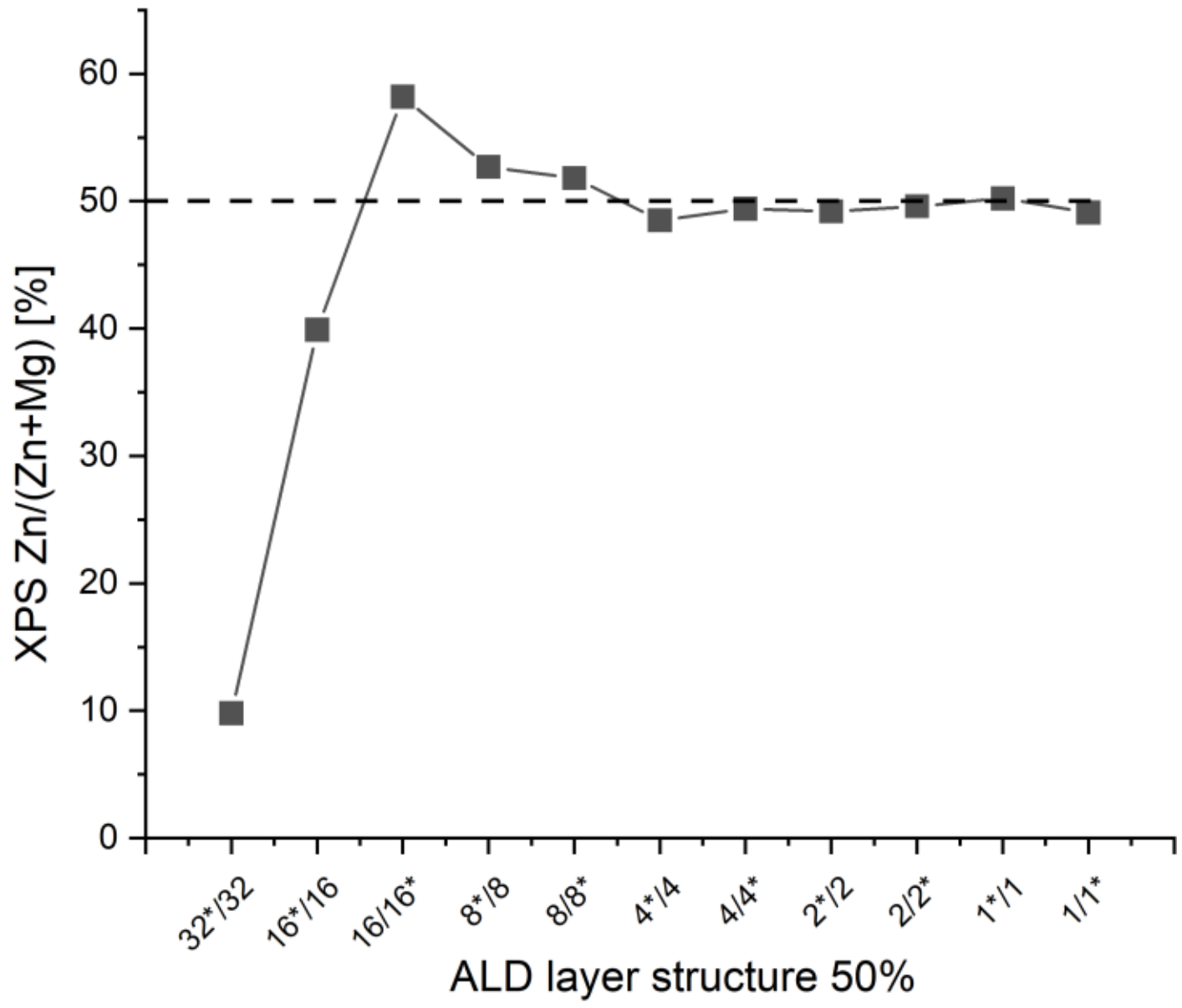

Figure 2: Graph showing Zn ratio of series 2 heterostructures after ion etching. The relative error bars for each measurement are combined in the thickness of the dots.

For heterostructures of the series 2 (Figure 2) with a number of successive ZnO and MgO cycles equal to or less than 8, the intrinsic Zn ratio remains very close to 50 ± 2%, in agreement with series 1 samples. For heterostructures with 16 or more successive ZnO and MgO cycles, the Zn ratio diverges greatly from 50%. This effect can be explained by the limited analysis depth of XPS, which is approximately 10 nm. As will be discussed later, these heterostructures consist of alternating ZnO and MgO layers. Consequently, for the thickest layers, the XPS probing depth is not large enough to average the composition over a sufficient number of layers and the measured Zn ratio becomes strongly influenced by the composition of the topmost surface sheet. As a result, the 16/16* sample, which is terminated by a pure 2.5nm thick ZnO layer, exhibits a measured Zn ratio of 58.2%, whereas the 16*/16 sample, whose surface corresponds to a pure 2.1 nm thick MgO layer, shows a significantly lower Zn ratio of 39.9%.

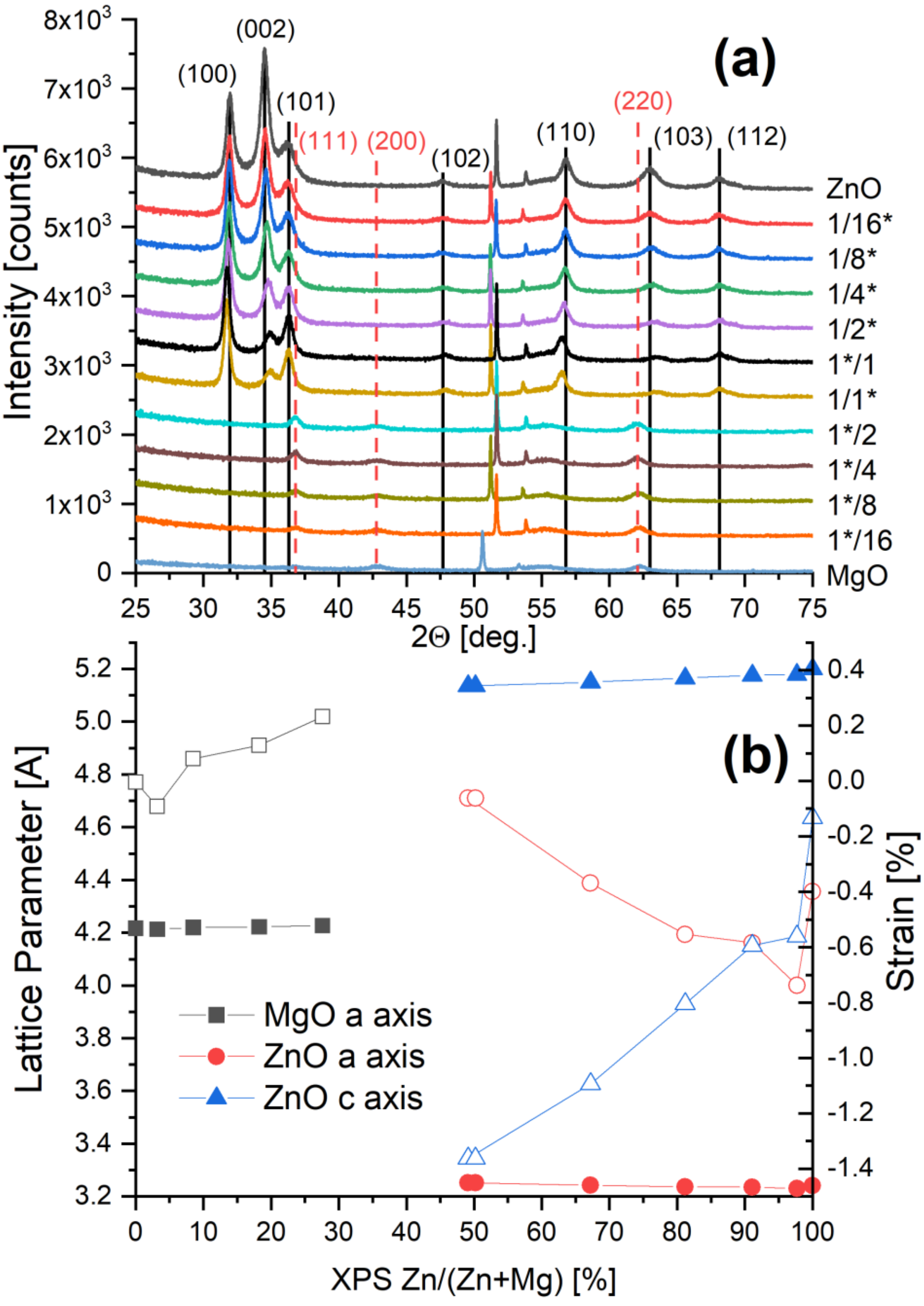

Figure 3: XRD diffractograms (a) and the associated lattice parameters and strains (b) for the ZnMgO heterostructures in series 1. The relative error bars for each measurement are combined in the thickness of the dots.

In order to further investigate the observed composition trend, we performed XRD and XRR analysis of the heterostructures.

**b) Heterostructure structure characterization**

The XRD diffractograms of the series 1 heterostructures are represented in Figure 3.a. They reveal that when the Zn ratio is greater than or equal to 50% (50% or more ZnO cycles), the resulting microstructure is the one of ZnO. These results are similar to those reported by Ting Xie et al. [31] under the same growth conditions. For the coatings 1/1*, 1*/1, 1/2*, 1/4*, 1/8*, 1/16* and pure ZnO, the diffractograms confirm that the microstructures are hexagonal (wurtzite) and the analysis performed using CrystalDiffract® showed that the samples exhibit gradual shift of preferred orientations: from (002) for the 1/16* heterostructure and pure ZnO in agreement with literature [29], toward (100) as the content of ZnO (or the number of successive ZnO cycles) decreases. Heterostructures containing a majority of MgO ALD cycles—namely 1*/2, 1*/4, 1*/8, 1*/16, as well as pure MgO—display the expected body-centered cubic (MgO-type) microstructure. Similarly to ZnO, the diffractograms also display a gradual shift of preferred orientation from (200) for pure MgO, in agreement with literature, to (111) orientation for mixed compositions.

The diffractograms also highlight a sharp transition in crystallinity, shifting abruptly from the wurtzite ZnO when ZnO ALD cycles dominate to cubic MgO structures when MgO cycles dominate. This behavior contrasts with other mixed heterostructure systems—such as NbTiN [33] or $ErYO_3$ [33]—where the lattice parameter typically evolves more gradually toward that of the pure compounds. This difference likely stems from the fact that in those systems the end-member materials (NbN/TiN and $Er_2O_3$/$Y_2O_3$) share the same (cubic) crystal structures, whereas ZnO and MgO in ZnMgO alloys possess very different crystalline structures.

Figure 3b presents the lattice parameters and associated strains for series 1 heterostructures and extracted from the diffractograms described in the supplemental informations. Heterostructures with a ZnO-type microstructure, the lattice parameter c decreases from 5.20 Å to 5.13 Å, while the parameter a slightly increases from 3.24 Å to 3.25 Å as the Zn ratio is reduced from 100% to 50%. These changes indicate that reducing the number of successive ZnO ALD cycles—and thus decreasing the ZnO layer thickness—results in greater compressive strain along the c-axis and lower compressive strain along the a-axis. By comparison, heterostructures with an MgO-type microstructure appears to be much less sensitive to composition variations with little strain.

We can conclude that the above structural evolutions for series 1 (i.e preferential orientation and tensile strain) indicate that the ZnO wurtzite hexagonal and MgO cubic structures tend to adapt to each other at maximal 1/1 (50%) mixing, following the pattern shown in Figure 4, where the hexagonal structure is favored, then relax towards those of the pure materials when the number of successive ALD cycles increases (for either ZnO or MgO). In turn this affect the ZnO growth rate observed in figure 2. Density Functional Theory (DFT) calculations would be needed to confirm this hypothesis.

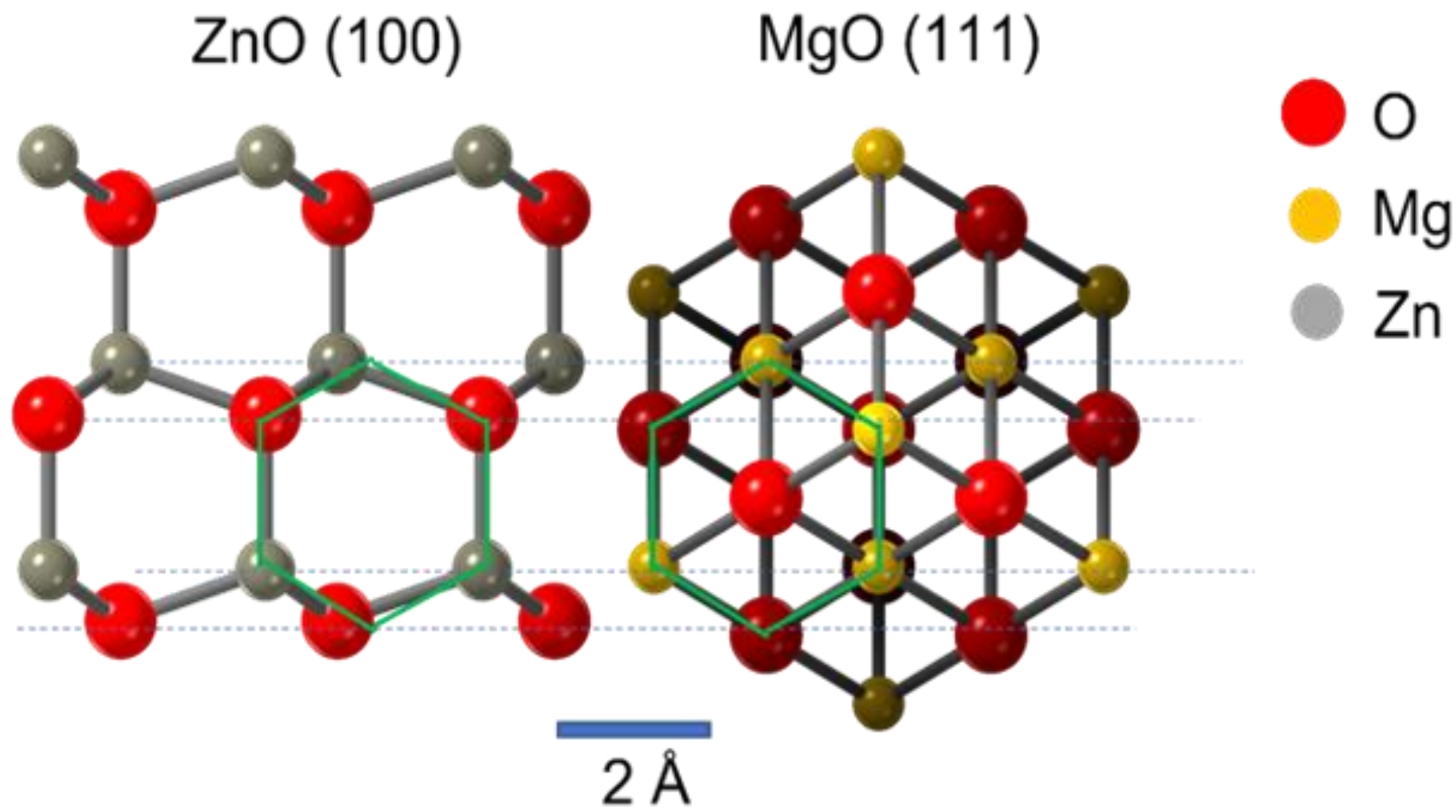

Figure 4: Representation of the adaptation between a hexagonal (ZnO) and cubic (MgO) structure for ZnO: MgO = 1:1, deduced from the series 1 diffractograms.

The fits of the XRR measurements are summarized in Figure 5. The average GPC, calculated by dividing the total film thickness by the total number of ALD cycles, follows the same trend as the multilayer crystalline structure previously analyzed: a GPC close to the MgO GPC ~ 1.2-1.3 A/cy in MgO rich films and sharp transition to a GPC close to the ZnO ~ 1.7-1.8 A/cy in the ZnO rich films. The theoretical density is calculated as:

$$\text{Theoretical density} = \text{Zn ratio} \times \rho_{\text{pure ZnO}} + \text{Mg ratio} \times \rho_{\text{pure MgO}} \quad (4)$$

Where :

- Zn ratio is calculated from equation (1),
- $\rho_{\text{pure ZnO}}$ is the pure ZnO coating density (200 cycles), measured by XRR = 5.55 g.cm-3,
- $\rho_{\text{pure MgO}}$ is the pure MgO coating density (250 cycles), measured by XRR = 3.33 g.cm-3,
- Mg ratio = 1 - Zn ratio.

Figure 5 reveals two distinct regimes in the composition dependence of film density. In the MgO-rich regime (0–28% Zn), the measured density exceeds the theoretical value, with an increasing deviation as the Zn fraction rises and the growth per cycle (GPC) decreases. As discussed above, increasing Zn incorporation favors MgO (111) texture. Experimentally the product of both the density and the GPC is constant (44 ± 2 ng/cm$^2$) and account for a surface density of atoms incorporated per unit area during each ALD cycle, governed by surface reaction probabilities and the density of reactive sites. As a results a denser film yields a smaller thickness increment per cycle. In the ZnO-rich regime (50–100% Zn), the density increases linearly with Zn content and approaches the theoretical value. Within this regime, both density and GPC exhibit only weak sensitivity to the incorporation of MgO layers. These results indicate that, in this system, density is primarily composition-driven, whereas GPC is dictated by microstructural character.

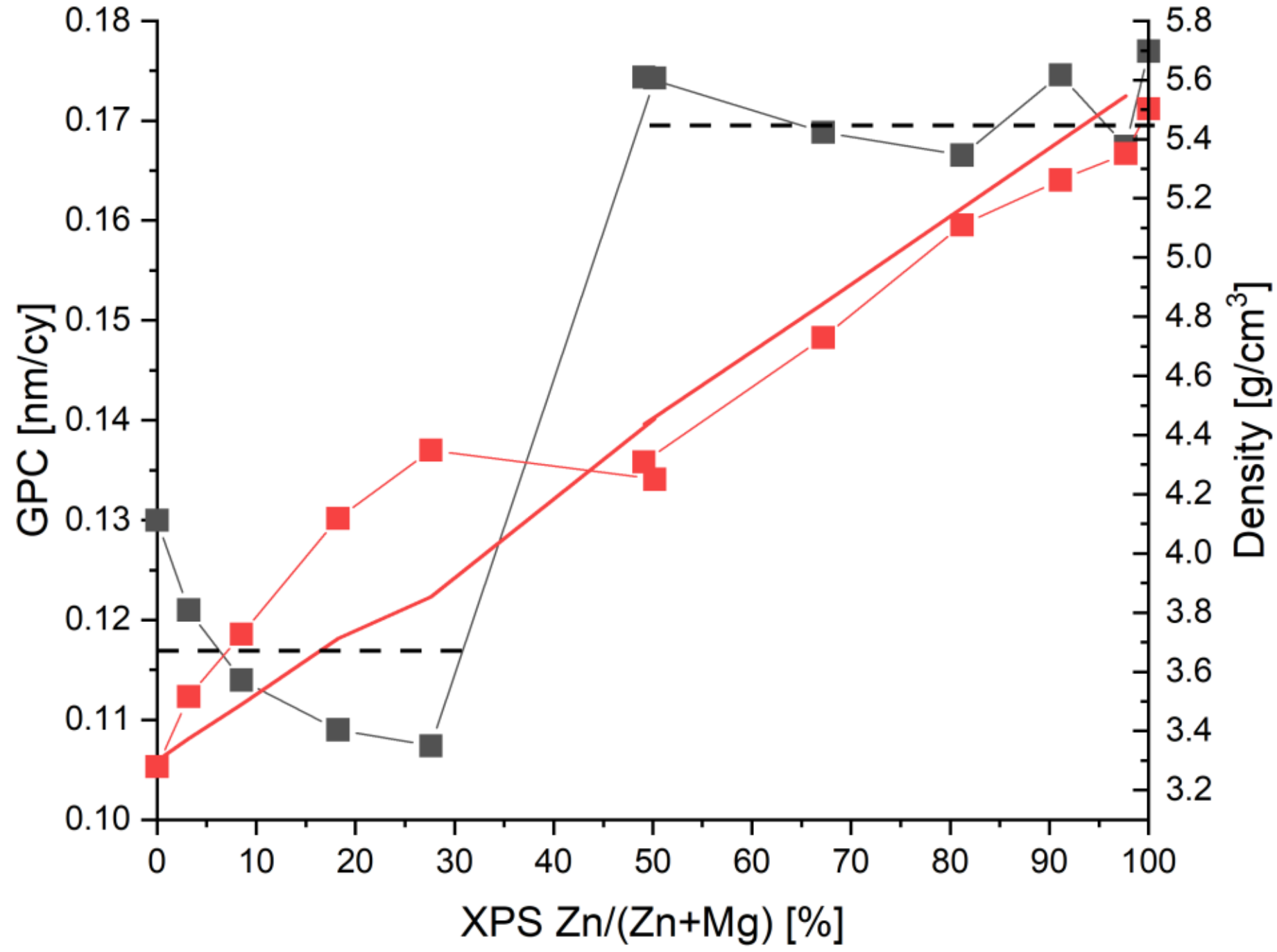

Figure 5: Averaged growth rate per cycle (GPC) in red and density in black as a function of Zn ratio in series 1 samples. The relative error bars for each measurement are combined in the thickness of the dots.

The series 2 heterostructures (Figure 6.a) exhibit ZnO-type hexagonal (wurtzite) microstructures with a preferred (100) orientation, consistent with the 1:1 (50%) sample from series 1. Heterostructures composed of eight or more successive ALD cycles of ZnO and MgO display, in addition to the characteristic ZnO peaks, low-intensity peaks corresponding to body-centered cubic MgO with a preferred (200) orientation, consistent with pur MgO films. It is also evident from the diffractograms that the growth order (ZnO or MgO first) does not influence the crystalline structure of the multilayers, as expected. XRR measurements (Figures 7.b and 7.c) further confirm the presence of well-defined ZnO and MgO layers when the number of successive ALD cycles reaches 8 or more.

Relative to series 1, increasing the number of MgO ALD cycles drives the MgO crystalline structure towards that of its bulk (200) phase. In contrast, for an equivalent number of ZnO cycles (for example, 1/2 versus 2/2 and 1/4 versus 4/4), the ZnO structure—particularly the (002) orientation—progressively departs from that of pure ZnO films as the MgO layer thickness increases. This trend indicates that ZnO crystallinity orientation is strongly governed by the underlying MgO layer. This influence is further evidenced by the progressive attenuation and eventual disappearance of ZnO diffraction peaks with increasing numbers of successive ALD cycles, despite the concomitant increase in overall thickness (Fig. 7b). Such behavior suggests a transition of ZnO nanolaminates towards an amorphous state as the MgO contribution increases and the MgO (200) orientation becomes dominant. In contrast with series 1, this may reflect a structural incompatibility, whereby hexagonal ZnO more readily accommodates a MgO hexagonal-like (111) arrangement than a cubic (200) orientation. At sufficiently large ZnO thicknesses (>32 cycles), a recovery towards bulk-like ZnO crystallinity is expected, irrespective of the MgO cycle number, although this regime lies beyond the scope of the present study.

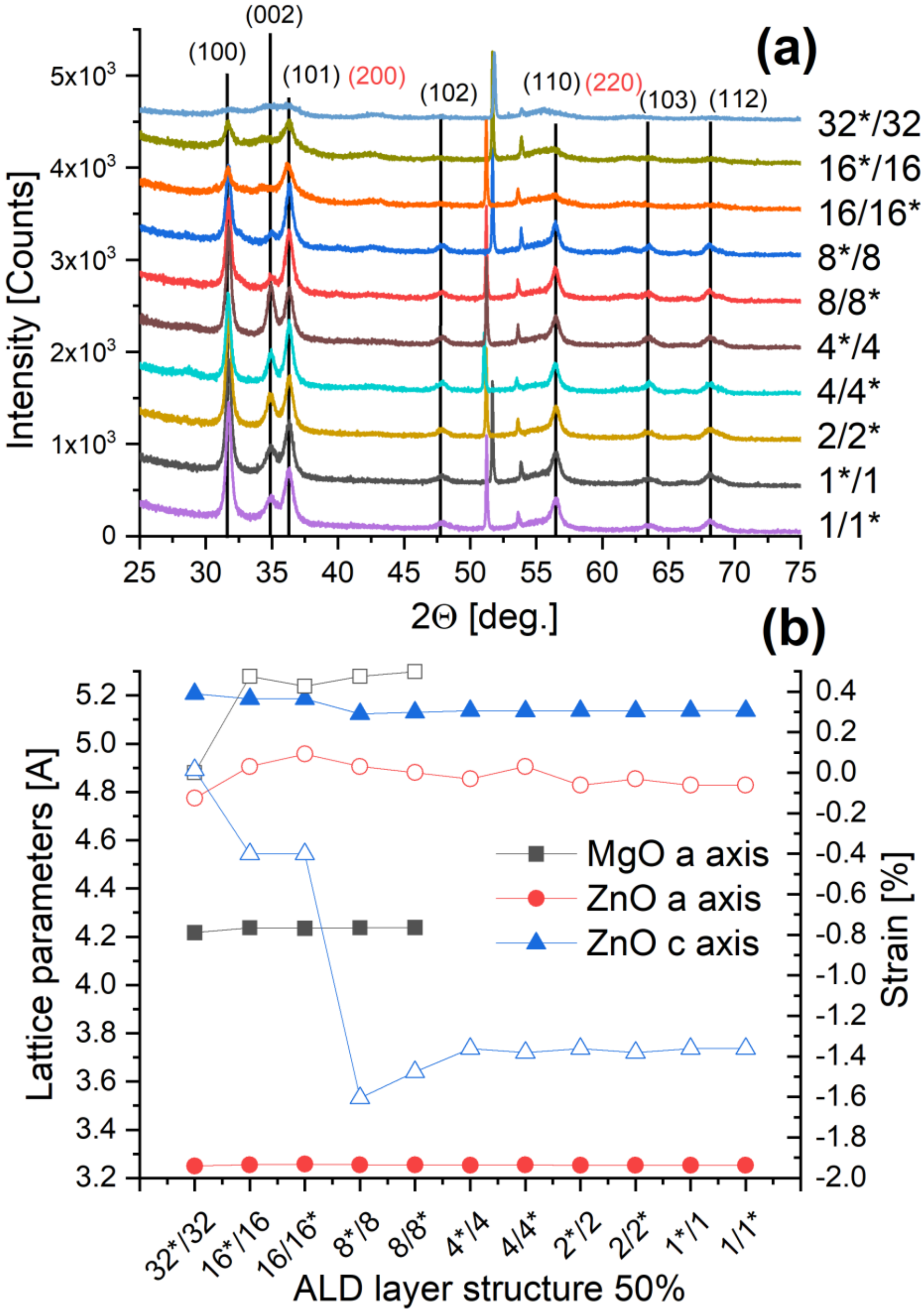

Figure 6: XRD diffractograms (a) and the associated lattice parameters and strains (b) for the ZnMgO heterostructures in series 2. The relative error bars for each measurement are combined in the thickness of the dots.

Analysis of the diffractograms (Fig. 6b) shows that the ZnO lattice parameter ***a*** remains nearly constant (3.25–3.26 Å), corresponding to low uniaxial tensile strains (0 ± 0.1%), independent of the number of successive MgO and ZnO ALD cycles. The evolution of the c axis parameters is quite different; for heterostructures comprising eight or fewer successive cycles, the c parameter remains stable (5.13–5.14 Å), indicating a constant and compressive strain. At higher cycle numbers (16 and 32), this compressive strain is significantly reduced and simultaneously the ZnO film crystallinity vanishes in contrast to series 1 samples.

The MgO lattice parameter ***a*** remains comparable to that of series 1, except for the 16/16* and 16*/16 samples, which exhibit reduced compressive strain. Notably, the 16/16*, 16*/16, 1/1*, and 1*/1 samples display distinct strain states, suggesting corresponding variations in their electrical and optical properties, as lattice strain is known to influence properties such as piezoelectric response, bandgap, and charge transport.

X-ray reflectivity measurements (Fig. 7) further indicate that, from eight successive ALD cycles onward, the films consist of well-defined alternating ZnO and MgO layers. In this regime, film density increases with the number of successive cycles, in contrast to structures with 1, 2, or 4 successive cycles. For these films, no evidence of distinctive ZnO and MgO layers were shown by XRR and XRD measurements, possibly due to a lack of sensitivity of the measurements. For heterostructures such as 8/8, 16/16, and 32/32, the densities of individual ZnO and MgO layers approach those of their respective bulk-like films (series 1) as layer thickness increases, consistent with enhanced structural ordering rather than amorphization.

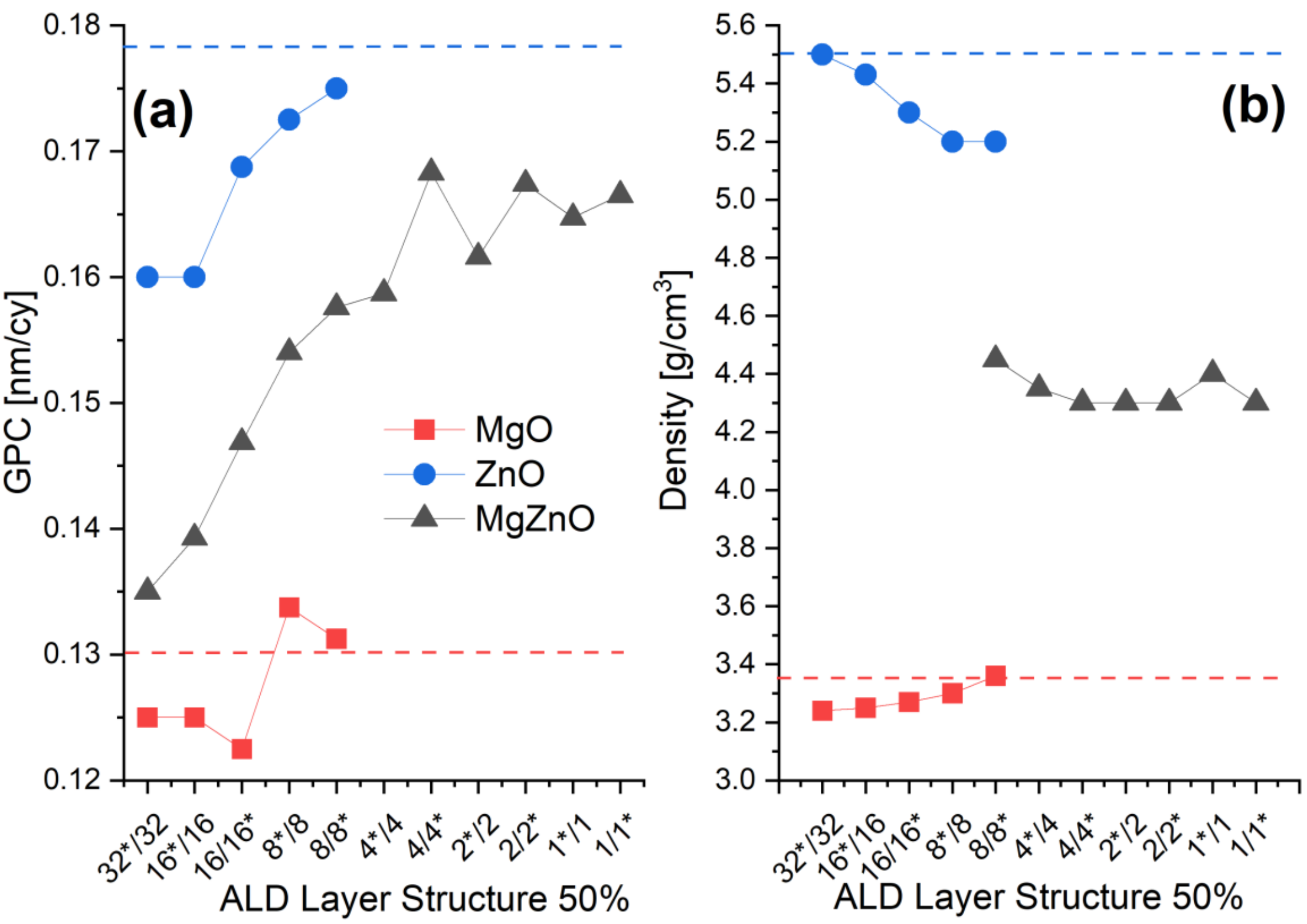

Figure 7: Thickness and density of series 2 heterostructures (a) ; Averaged thickness and density of ZnO and MgO layers in series 2 samples (b) ; GPC of heterostructures in series 2 (c). The relative error bars for each measurement are combined in the thickness of the dots.

By contrast, heterostructures formed with 1, 2, or 4 successive cycles exhibit similar densities, GPC values, and microstructures, indicating comparable ALD growth behavior and explaining the absence of discernible microscopic structural differences in this regime.

### c) Electrical conductivity analyses

The electrical conductivities of heterostructures from series 1 and 2 are presented in Fig. 8. For series 1, conductivity is shown as a function of Zn content. Samples not displayed exhibited conductivities below the detection limit of the measurement setup (≈0.01 $\Omega^{-1}$ $m^{-1}$). Despite this limitation, the data reveal clear trends linking electrical conductivity to both the chemical composition and the structural characteristics of the heterostructures.

The conductivity decreases exponentially (red curve in Fig 8a)) within the heterostructures exhibiting a ZnO-type microstructure. Notably, pure ZnO coatings display a conductivity of $1.5 \times 10^4$ $\Omega^{-1}$ $m^{-1}$, approximately twice that reported for comparable ALD-grown films under similar conditions (DEZ/$H_2O$, 150 °C, thickness 30–100 nm). The conductivities of the 16*/16, 16/16*, 8*/8 and 8/8* coatings are 178, 86, 18 and 15 $\Omega^{-1}$ $m^{-1}$, respectively. Figure 9b shows that, in series 2, conductivity decreases with the number of successive ZnO and MgO ALD cycles. The heterostructures represented consist of well-defined ZnO and MgO layers. As shown in Fig. 6c, reducing the number of successive cycles decreases both the thickness and density of the individual layers, particularly ZnO, resulting in reduced conductivity.

These results indicate that the presence of distinct ZnO and MgO layers enhances conductivity relative to heterostructures with similar Zn ratios but without layered structuring. Charge transport is dominated by these layers, whose conductivity increases with thickness and density, both governed by the number of successive ZnO cycles. Accordingly, layered heterostructures with a Zn ratio of 50% (figure 8 b)) exhibit higher conductivity than the 67% Zn sample in Fig. 8a), demonstrating that macroscopic structuring provides an additional degree of control beyond composition alone.

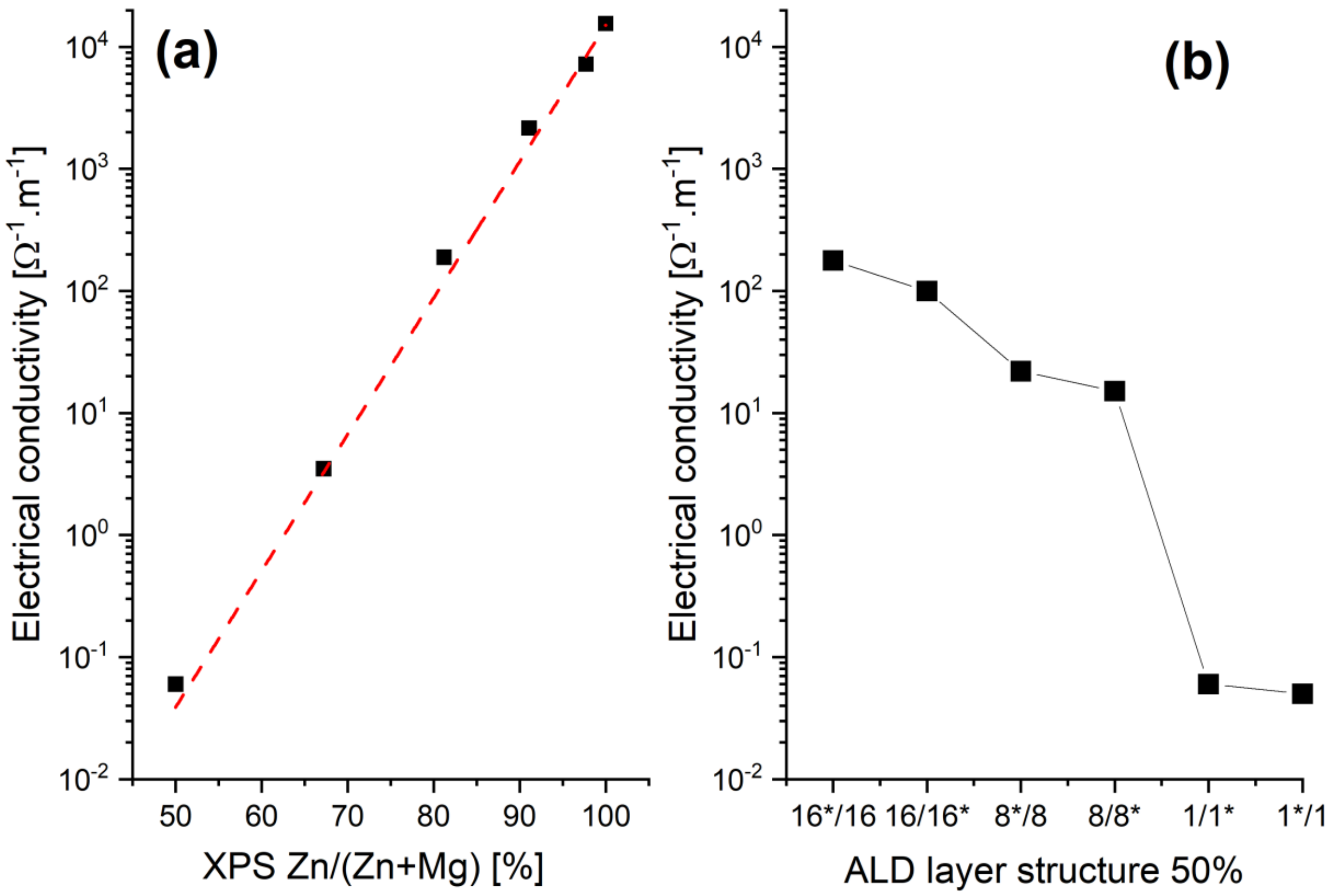

Figure 8: Electrical conductivity measured on the heterostructures of series 1 (a) as a function of their Zn ratio

**b) TEEY analyses**

TEEY measurement obtain from the different sample are acquired as function of incident electron energy. As shown in figure 9, presenting the TEEY of a pure ZnO film, the TEEY first increases until reaching a maximum of TEEY ($TEEY_{max}$) at electron energy $E_{max}$ then decreases.

It can be observed in figure 9 that in the as received condition, the variations in TEEY are smaller than those measured after etching. This behavior is attributed to the presence of a surface contamination layer composed of adsorbed hydrocarbons and hydroxides. Such a layer affects the production and transport of excited electrons, as well as the material work function, leading to TEEY values that differ from those of the intrinsic coating. A similar trend is also observed for every series 1 and 2 samples respectively shown in Figure 10.a and 10.b.

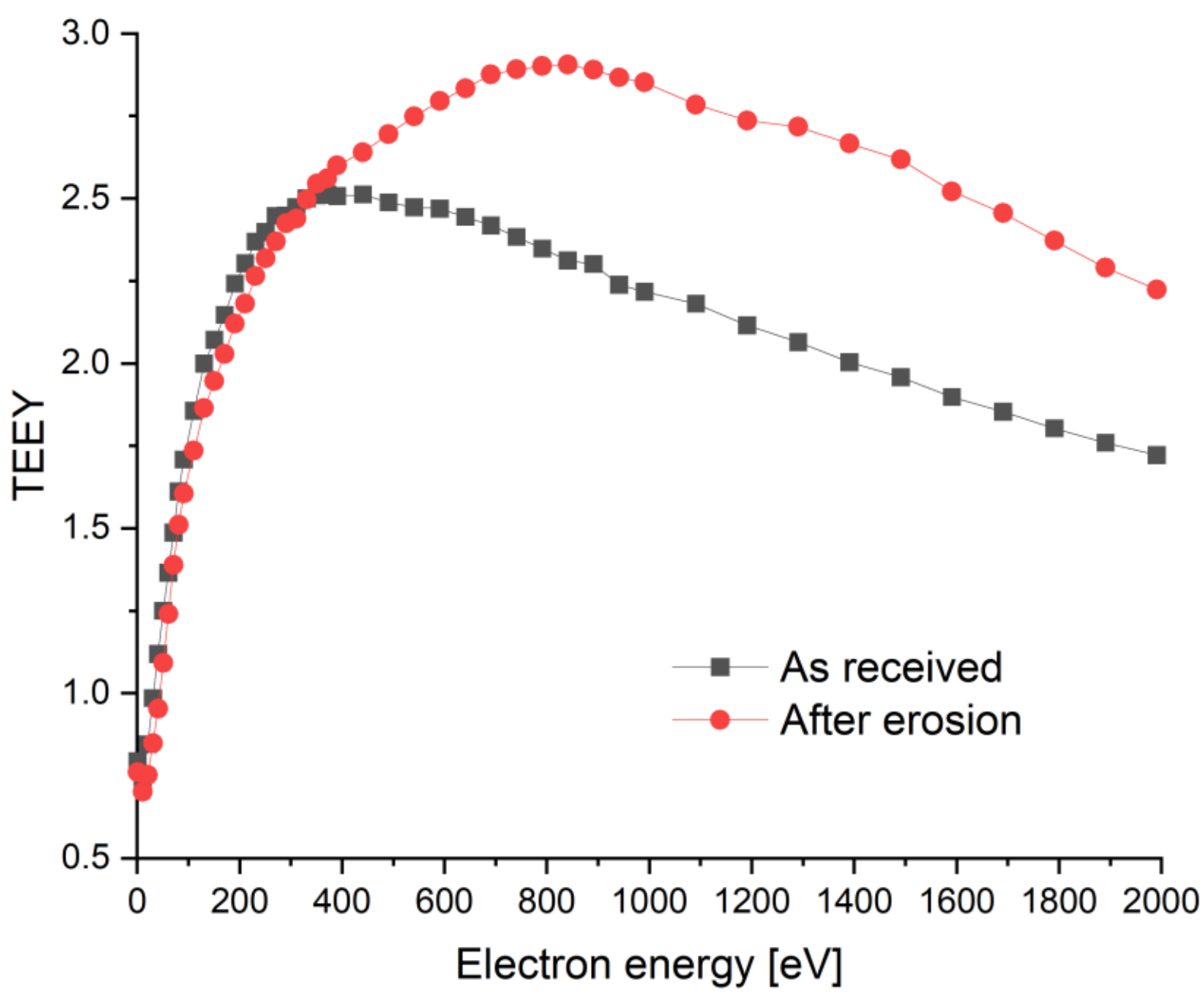

Figure 9: TEEY curves of ZnO ALD thin film measured as received and after ion etching. The relative error bars for each measurement are combined in the thickness of the dots.

Figure 10.a presents the maximum measured TEEY for series 1 heterostructures as a function of their Zn ratio. The TEEY of the etched surfaces (intrinsic maximum of TEEY) is seen to increase as the Zn ratio decreases from 100% to 50%. However, when the Zn ratio falls between 27% and 0%, the intrinsic maximum TEEY remains nearly constant, ranging from 6.8 to 6.1, regardless of composition. This plateau is likely due to charge accumulation during each electron pulse in the measurement process, which introduces artifacts in the extracted TEEY values. As a result, the reported maximum TEEY values in the 27–0% Zn range should be regarded as lower-bound estimates rather than exact values. Nevertheless, the measured maximum TEEY for pure MgO remains in excellent agreement with the literature (6.1 measured vs. 6.2 reported).

Notably, the TEEY shows a marked increase when the Zn ratio decreases from 50% to 27%, rising sharply from 3.9 to 6.8, whereas it increases by only about 1 unit between 100% and 50% Zn. This pronounced jump occurs at the same time as the structural transition from a ZnO-type microstructure to an MgO-type microstructure, as illustrated in Figure 3. This observation highlights the dominant influence of MgO in these films, since the measured TEEY becomes comparable to that of pure MgO. Overall, these results indicate that both the chemical composition and the resulting microstructure—which is closely linked to the Zn ratio—play a key role in determining the intrinsic TEEY behavior of the heterostructures.

Figure 10.b indicates that, after etching, the maximum TEEY measured for series 2 heterostructures produced with 1 to 4 successive ALD cycles of ZnO and MgO changes only slightly (from 3.5 to 4.1) and remains essentially independent of the heterostructure architecture (See table 1). This agrees with the results presented in Figures 6 and 7, which showed that these samples share very similar characteristics in terms of density, growth-per-cycle (GPC), lattice strain, and microstructure.

In contrast, when the heterostructures are synthesized using 16 successive cycles or more, a clear increase in maximum TEEY becomes visible and depends directly on the number of cycles: the larger the number of consecutive cycles, the higher the TEEY. This demonstrates that the macrostructuring of the heterostructures significantly influences their TEEY behavior.
Interestingly, the intrinsic maximum TEEY values measured for the 16/16* and 16/16* heterostructures are almost identical, even though the thickness of the top surface layer (ZnO or MgO) exceeds 2 nm. Since TEEY is strongly dependent on surface chemistry (which controls electron emission, transport, and the work function, as highlighted by the effect of carbon contamination in the as-received measurements) this lack of sensitivity to the surface layer composition is unexpected and not explained at this time.
For the series 2 samples, no clear relationship was found between the maximum TEEY (whether as received, after conditioning, or after etching) and the measured crystallite size, density, lattice parameters, or uniaxial strain. A similar conclusion applies to the series 1 samples, except in the case of heterostructures exhibiting a ZnO-type microstructure, where uniaxial strain appears to play a role.

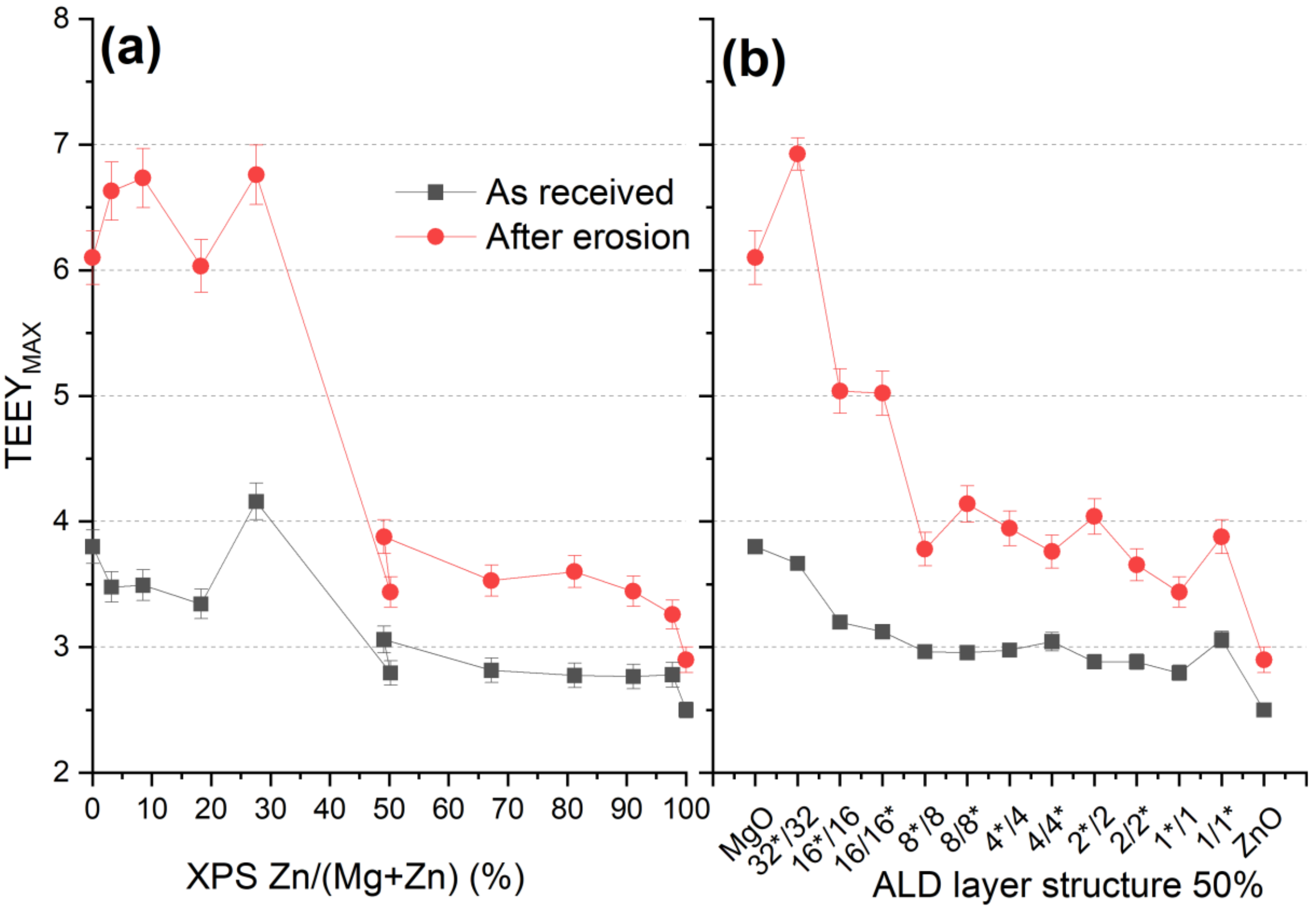

Figure 10: Maximum TEEY for series 1 (a) and 2 (b) measured as received and after ion erosion.

As shown in Figure 11.a for strain along the c-axis the maximum TEEY tends to increase with higher compressive strain along the c-axis.
These trends may be attributed to strain-induced changes in the ZnO band gap. Previous studies have shown that the band gap of pure monocrystalline ZnO increases under uniaxial strain—whether tensile or compressive—applied along the c-axis [34]. A comparable effect has also been reported for MgO; however, because MgO has a band gap more than twice that of ZnO, the relative change in MgO is expected to be much smaller. Secondary and backscattered electrons with energies superior to the band gap energy plus electronic affinity can only interact with phonon thus have a much greater escape depth than lower energy electrons [35]. So, ZnO layers stressed along c-axis may have wider band gap hence a higher TEEY than those with lower c-axis stresses.
It is worth recalling that monocrystalline ZnO is a semiconductor with a room-temperature optical band gap of approximately 3.37 eV. In contrast, ZnO deposited by ALD, either as a pure film or within mixed heterostructures, cannot be considered fully intrinsic due to the presence of defects introduced during growth (Hydroxyls groups,

O vacancies…). Consequently, its band gap is typically lower, reported at 3.30 eV [36, 37] and decreasing when the number of ZnO ALD cycles increases [37]. Under these conditions, any strain-driven band gap increase would represent a proportionally larger relative variation than in monocrystalline ZnO. This could explain why TEEY variations are observed as a function of ZnO microstructural strain, whereas a similar dependence is not clearly seen for MgO strain.

Figure 12 presents the maximum TEEY measured after etching for the samples whose electrical conductivity could be determined, together with their corresponding conductivity values. Two key outcomes clearly emerge from this figure.

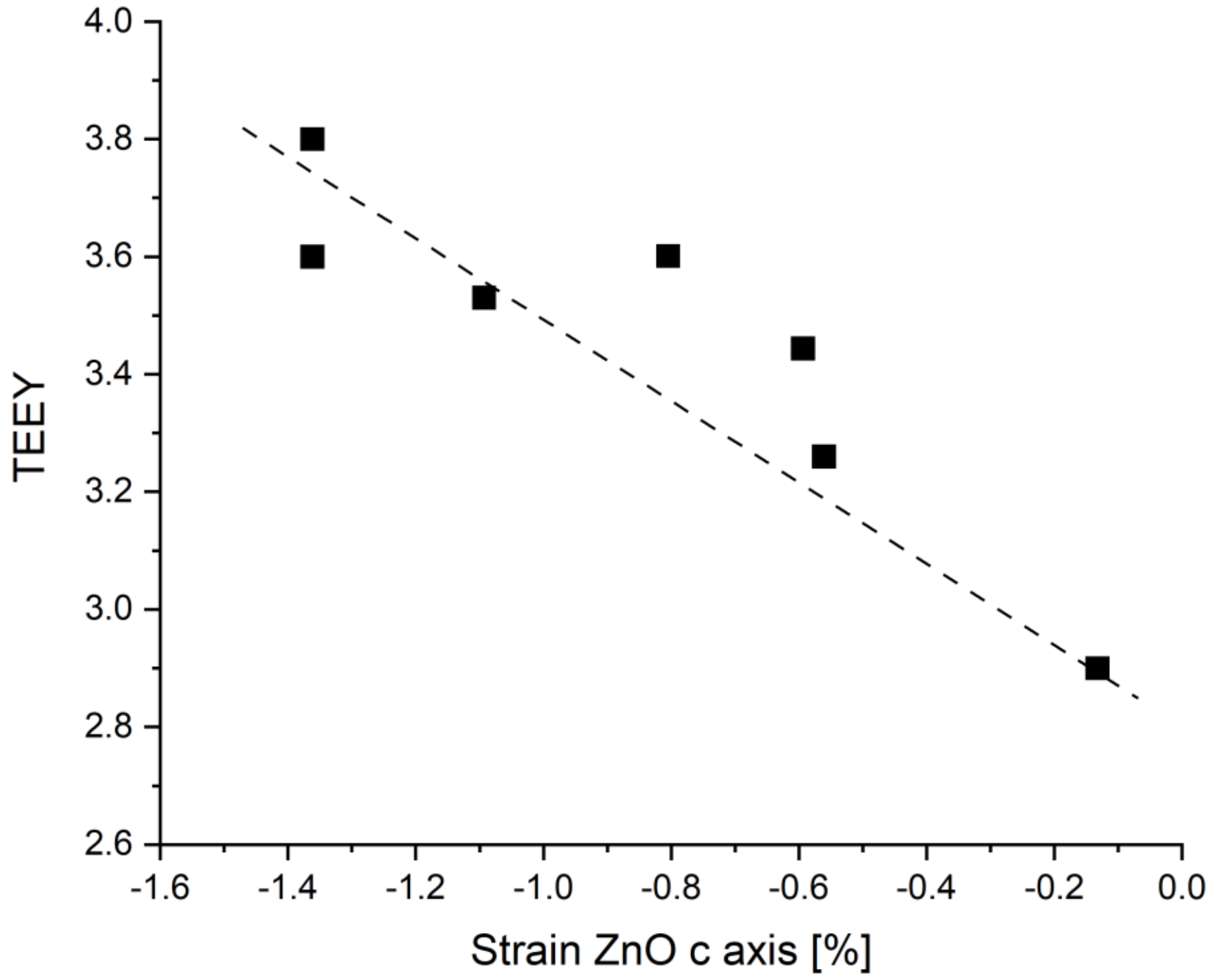

Figure 11: Maximum TEEY of series 1 heterostructures for different surface states as a function of uniaxial stresses along the c-axis (a) and the a- or b-axis (b).

As the Zn fraction decreases from 100% to 67%, the electrical conductivity drops exponentially, whereas the maximum total electron emission yield (TEEY) after etching increases only marginally, from ~3 to ~3.5. Notably, for the 1/1* and 1*/1 heterostructures (50% Zn), the maximum TEEY remains relatively low (~3.8), despite an electrical conductivity of ~0.06 $\Omega^{-1}$ $m^{-1}$ (Fig. 8). This demonstrates that heterostructures can be engineered to combine low TEEY with low electrical conductivity. In this composition range (100–50% Zn), the microstructure remains ZnO-like and electrical conductivity decreases by nearly six orders of magnitude, whereas TEEY increases by only ~33%. This indicates that electrical conductivity and the carrier density and mean free path are significantly more sensitive to thickness variation than TEEY within this regime. Direct measurements of carrier density and mobility would provide further insight into their respective roles in governing both electron emission and transport properties.

For the 16/16*, 16*/16, 8/8*, and 8*/8 heterostructures, increasing the number of successive ALD cycles from 8 to 16—thereby producing thicker and denser individual layers—leads to a concurrent increase in both the maximum TEEY after etching and the electrical conductivity. This behaviour highlights the possibility of engineering coatings that combine high TEEY with high conductivity. We attribute this trend to the layered geometry of the heterostructures and the anisotropy of the measurements; the TEEY is probed perpendicular to the ALD layers whereas the electrical conductivity measure both perpendicular and parallel to these ALD grown ZnO and MgO layers. During TEEY measurements, emitted electrons may traverse multiple layers with heterogeneous properties; however, this effect diminishes as the thickness of individual layers increases. In the limiting case of the 32/32* heterostructure, the shallow escape depth of secondary electrons implies that emission is dominated by the surface layer alone. This is consistent with the high TEEY (~7) observed for the MgO-terminated 32*/32 sample. Electrical conductivity initially increases with layer thickness, but becomes unmeasurable for the thickest 32*/32 sample, showing that the insulating top most MgO layer is thick enough to prevent the current path to the underneath ZnO layers. There is therefore an optimal layer thickness—here on the order of 2 nm or 16 cycle of MgO—that provide the optimal compromise between charge transport and TEEY. These results indicate that careful control of ZnO and MgO layer thicknesses enables simultaneous optimization of electrical conductivity and electron emission properties.

Overall, these findings confirm that two-material heterostructures can provide access to combinations of TEEY and electrical conductivity that cannot be achieved using single-material coatings alone.

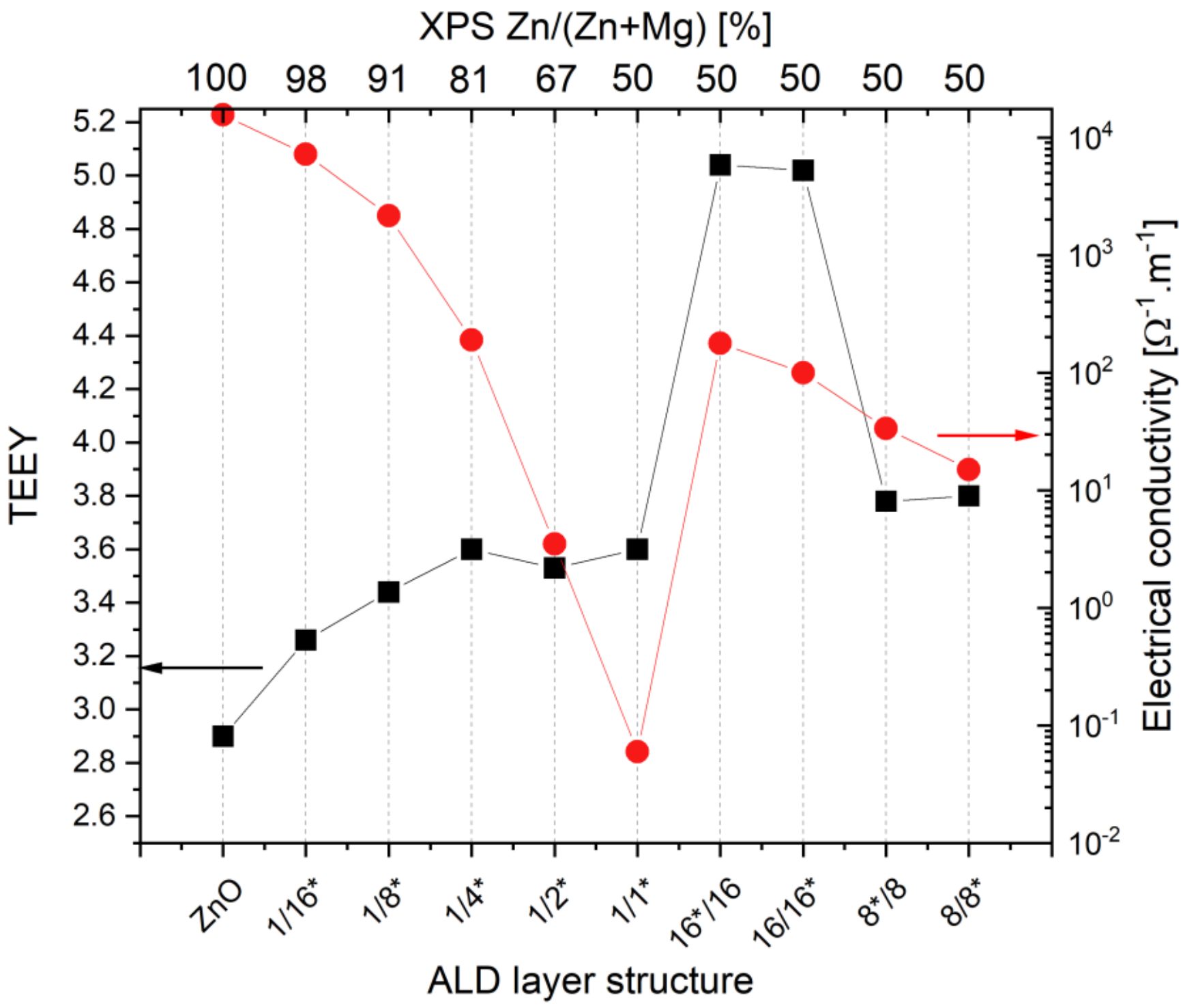

Figure 12: Electrical conductivity and maximum TEEY after ionic erosion of coatings for which electrical conductivity could be obtained.

## Conclusion

We have systematically investigated the microstructural, structural and chemical evolution of ZnO–MgO heterostructures as a function of ALD cycle sequencing across a library of 18 samples. By varying the succession of ZnO and MgO deposition cycles, we demonstrate independent control over composition, structure and microstructure, enabling precise tuning of functional electronic properties.

Measurements of total electron emission yield (TEEY) under different conditioning states, together with electrical conductivity for the most conductive samples, reveal two distinct design regimes. These heterostructures can be engineered to simultaneously maximize or minimize both TEEY and electrical conductivity—property combinations that are not accessible in single-phase materials. In particular, coatings combining low TEEY and low conductivity are promising for superconducting radio-frequency cavities, whereas high-TEEY, high-conductivity systems may benefit applications such as microchannel plates.

These results establish heterostructuring as a versatile strategy for decoupling and tailoring electron emission and transport properties. Future work will focus on exploring alternative material pairs to broaden the accessible property space, and developing predictive models linking fabrication parameters to functional performance.

## Acknowledgements

This work was supported by Internal CEA Ph.D. funding (CFR) and by the European Union's Horizon 2020 Research and Innovation programme under Grant Agreements No 101004730 (IFAST) and No 101057511 (EURO-LABS).

## Supplementary Materials

The cell parameter values for body-centered cubic MgO and hexagonal ZnO microstructures, obtained by calculation and with CrystalDiffract@. The uniaxial strain was calculated as follows:

$$\varepsilon = \frac{x - x_0}{x_0} \times 100 \quad (3)$$

Where:

- ε is the uniaxial strain along the considered axis, in percent,
- x is the calculated lattice parameter,
- $x_0$ is the reference lattice parameter at room temperature and zero stress.

For ZnO, the reference for parameters a and c are respectively 3.2506 Å and 5.2057 Å [39]. For MgO, the reference for parameter a is 4.2506 Å [40].